\documentclass[%
reprint,
amsmath,amssymb,
aps,pra]{revtex4-1}

\usepackage{color}
\usepackage{graphicx}
\usepackage{dcolumn}
\usepackage{bm}
\usepackage{footmisc}
\usepackage{mathtools}
\usepackage{framed}
\usepackage{mathrsfs}
\usepackage{lipsum}

\def\d{{\rm d}}

\newcommand{\Span}{{\mathsf{Span}}}

\def\>{\rangle}
\def\<{\langle}

\newcommand{\map}[1]{\mathcal{#1}}

\begin{document}


\title{Heisenberg-limited  metrology with   coherent  control on the probes' configuration}

\author{Giulio Chiribella$^{1,2,3}$}\email{giulio@cs.hku.hk}
\author{Xiaobin Zhao$^{1}$}
\affiliation{$^1$  QICI Quantum Information and Computation Initiative, Department of Computer Science, The University of Hong Kong,
Pokfulam Road, Hong Kong 999077, China}
\affiliation{$^2$ Department of Computer Science, University of Oxford, Parks Road, Oxford OX1 3QD,  United Kingdom}
\affiliation{$^3$ Perimeter Institute for Theoretical Physics, Caroline Street, Waterloo, Ontario N2L 2Y5, Canada}

\nopagebreak

\begin{abstract}
A central feature  of quantum metrology is the possibility of  Heisenberg scaling, 
a quadratic improvement over the limits of classical statistics.
 This scaling, however,  is notoriously fragile to noise. While for some noise types  it can be restored through error correction, for other important types, such as dephasing, the Heisenberg scaling appears to be irremediably lost. 
 %
 Here we show that this limitation can sometimes be  lifted if  the experimenter has the ability to 
  probe physical processes in a coherent superposition of alternative configurations.
    As a concrete example, we consider the problem of phase estimation in the presence of  a random phase kick, which in normal conditions is known to prevent the Heisenberg scaling. We provide a parallel protocol that achieves Heisenberg scaling with respect to the probes' energy, as well as a sequential protocol that achieves Heisenberg scaling with respect to the total probing time.  
 In addition, we show that Heisenberg scaling can also be achieved for  frequency estimation in the presence of continuous-time dephasing noise,  by combining the superposition of paths with fast control operations.  
\end{abstract}

\maketitle

Quantum metrology, the precise estimation of physical parameters aided by quantum resources, promises striking  enhancements with respect to the limits of classical statistics \cite{giovannetti2004quantum,giovannetti2006quantum,giovannetti2011advances}.  The best known example  is the Heisenberg  scaling, corresponding to a root mean square error with inverse linear scaling $1/N$ when an unknown physical process is probed by $N$ entangled particles \cite{bollinger1996optimal,walther2004broglie,afek2010high},  or by a single particle in a sequence of $N$  time steps \cite{higgins2007entanglement,chen2018heisenberg,chen2018achieving}.  The inverse linear scaling $1/N$ amounts to a quadratic improvement over the standard quantum limit $1/\sqrt N$, corresponding to a classical statistics over  $N$ repeated experiments.

The Heisenberg scaling was originally derived for the estimation of  noiseless quantum processes \cite{heitler1954quantum,holland1993interferometric,braunstein1994statistical}. 
Later, it was found out to be   extremely fragile to noise \cite{huelga1997improvement,fujiwara2008fibre,demkowicz2012elusive}.
   This finding stimulated an extensive search for methods to counteract noise in quantum metrology \cite{smirne2016ultimate,sekatski2017quantum,zhou2018achieving,albarelli2018restoring}.   For certain types of noise, it was found that the Heisenberg limit can be restored by error correction \cite{zhou2018achieving}, weak measurements \cite{albarelli2018restoring}, and fast quantum control \cite{sekatski2017quantum}.  However, some important types of noise have so far resisted all attempts.  The prototype of such resistant noise types is  dephasing \cite{gardiner1991quantum}, corresponding to random fluctuations with  the same generator as the signal.   While the Heisenberg scaling can be restored for some particular models of correlated dephasing \cite{dorner2012quantum,jeske2014quantum},      all the existing methods give rise to an error  scaling worse than the Heisenberg scaling  in the standard setting involving  uncorrelated dephasing  \cite{huelga1997improvement,escher2011general,demkowicz2012elusive,sekatski2017quantum,zhou2021asymptotic}. Intermediate scalings between the Heisenberg and standard quantum limit  have been achieved using error correction \cite{kessler2014quantum,arrad2014increasing,dur2014improved,zhou2018achieving} or  nondemolition measurements \cite{albarelli2018restoring,rossi2020noisy},  but as long as all the probes experience uncorrelated dephasing processes,  the Heisenberg scaling remained so far unattainable.

\begin{figure}
\centering
\includegraphics[width=0.5\textwidth,trim=4 4 4 4,clip]{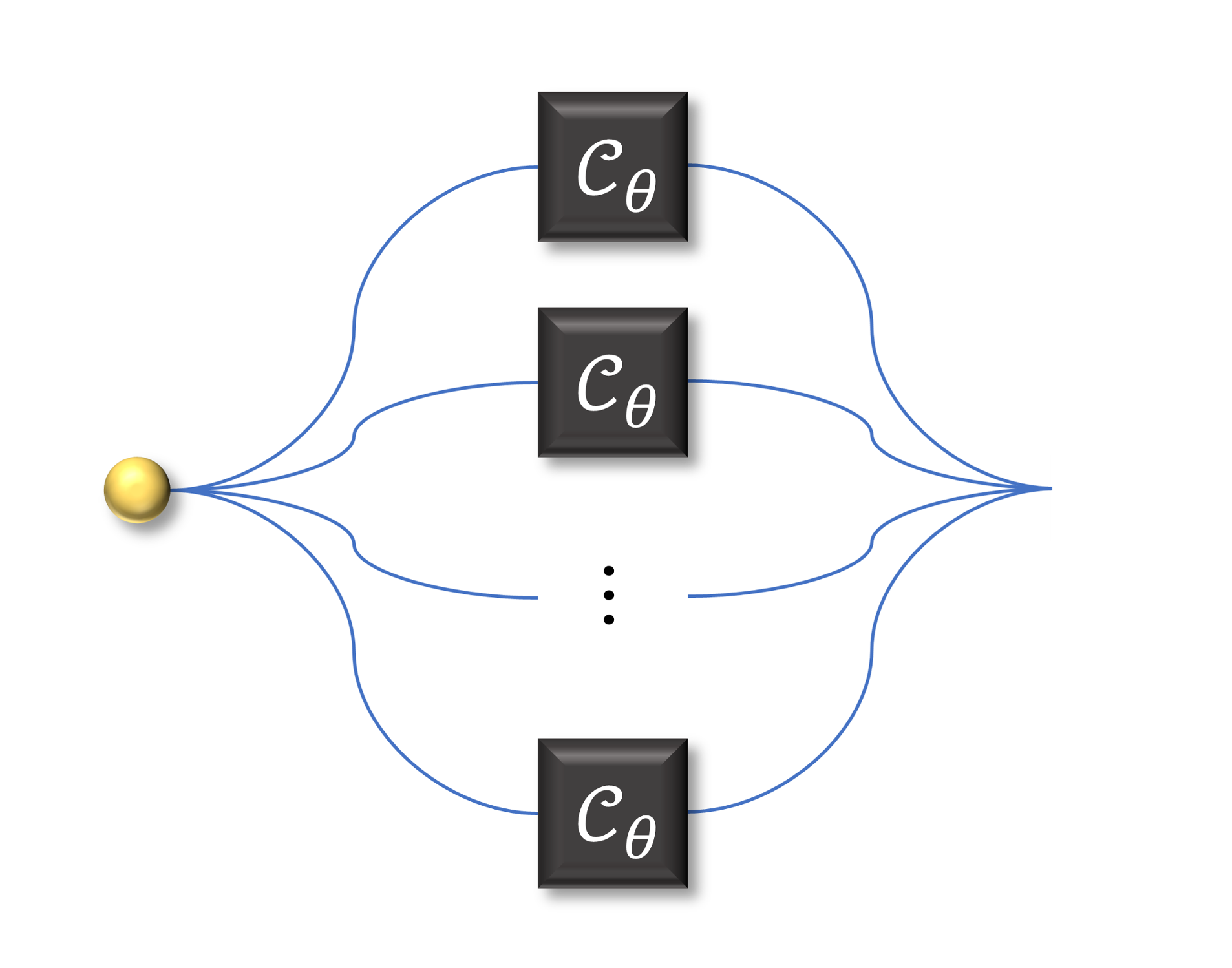}
{\caption{\label{fig:alternativepaths}   {\bf Single-particle probe in a coherent superposition of trajectories.}  A single quantum particle (in yellow) is used to probe an unknown  process depending on a parameter $\theta$.    The particle is sent through  $M$ alternative trajectories, each traversing an independinstance of the unknown process. In the end the trajectories are recombined and an interferometric measurement is performed.  This basic architecture can be used as a  building block for protocols using $N$ entangled probes, and for sequential protocols using a single probe in $N$ time steps. }}
\end{figure}

In this paper, we show that  the Heisenberg scaling   can  sometimes be achieved even in the presence of uncorrelated dephasing noise, provided that the experimenter has the ability to probe quantum processes in a coherent superposition  of multiple configurations \cite{aharonov1990superpositions,oi2003interference,aaberg2004subspace,gisin2005error,abbott2020communication,chiribella2019quantum,dong2019controlled,wechs2021quantum,vanrietvelde2021universal}.    
  We consider the  standard phase estimation problem where an unknown phase  $\theta$ is imprinted on the probes by  multiple uncorrelated instances of a noisy process $\map C_\theta$, and we explore  setups where each probe is sent through $M$ alternative trajectories, each leading to an  independent instance of the unknown  process $\map C_\theta$,  as illustrated in Figure \ref{fig:alternativepaths}.     We then use this architecture  as a building block for parallel protocols using $N$ entangled probes, and for sequential protocols using a single probe in $N$ time steps. 
Remarkably, we find out that the Heisenberg limit  can be restored, both in terms of  number of probes/total energy (for parallel protocols) and in terms of the number of time steps (for sequential protocols), when the process $\map C_\theta$ corresponds to a random phase kick, which shifts the phase either by $\theta$ or by $\theta +  \delta_0$, where $\delta_0$ is a fixed offset.  
  Building on these results, we show that the Heisenberg limit can also be  achieved for a physically relevant model of continuous-time Markovian dephasing \cite{gardiner1991quantum}, allowing an accurate frequency  estimation with an error decreasing quadratically with the total probing time.

Our results provide a new application of the technology  of coherent control over multiple trajectories, which has recently attracted increasing interest due to  its potential for  quantum communication   \cite{gisin2005error,abbott2020communication,chiribella2019quantum,loizeau2020channel,kristjansson2021witnessing}.  Superpositions of trajectories have been demonstrated experimentally  \cite{lamoureux2005experimental, rubino2021experimental} (see also 
\cite{goswami2018indefinite,wei2019experimental,guo2020experimental,rubino2021experimental,goswami2020experiments}
for related experiments using the superposition of trajectories to investigate indefinite causal order). Applications  of the superposition of trajectories to quantum metrology have been considered more recently \cite{chapeau2021quantum}  (see also  \cite{mukhopadhyay2018superposition,zhao2020quantum,chapeau2021noisy} for the study of quantum metrology with indefinite causal order). To the best of our knowledge, however, the fundamental question of  whether the superposition of trajectories could restore the Heisenberg  scaling in noisy quantum metrology  has remained unaddressed in the previous works. 
The results of the present paper answer this question in the affirmative, showing that  the Heisenberg scaling can be restored in the presence of a random phase kick,  a type of noise that is known to prevent the the Heisenberg scaling in all scenarios where the probes' configuration is fixed \cite{huelga1997improvement,escher2011general,demkowicz2012elusive,sekatski2017quantum,zhou2021asymptotic}.  


 The essential ingredient in all our protocols is coherence  in the path degree of freedom of single particles.  To achieve the Heisenberg scaling $1/N$, we find that the number of paths  $M$  needs to grow linearly with $N$. This means that the unknown process needs to be potentially available for probing in $O(N^2)$   configurations.   
While this still gives a  standard quantum limit in terms of the number of configurations, it is important to stress that the {\em potential} availability of a device in different configurations  is a much  weaker  resource than its actual use on a probe particle.  For example,     a photonic setup using  $N$  photons has the same energy independently on the number of paths on which the photons are routed.  Thanks to this fact, the superposition of paths offers a promising way to reach  high  precision while  employing a limited amount of energy.   Moreover,  putting a photon on a coherent superposition of paths only requires a sequence of beamsplitters, and  scaling up the number of paths   is generally easier than scaling up the number of photons in a multipartite entangled state.    Finally, it is also worth mentioning that our protocols do not require entanglement of  the $N00N$ type \cite{dowling2008quantum}: for our parallel protocols, we will only  use polarization entangled GHZ states  \cite{greenberger1989going}, which are comparatively easier to produce experimentally (see e.g. \cite{zhong201812} for a recent experiment with $N=12$).

\medskip




\emph{Phase estimation with superposition of paths.}--    Consider the estimation of a phase shift $\theta$ acting on a single qubit, for example corresponding to the polarization of a single photon.  In the ideal scenario, the parameter $\theta$ is imprinted on the system by a unitary gate $U_\theta:=\exp(-i\theta Z/2)$, with  $Z= \begin{pmatrix} 1  &  {\phantom -}0 \\  0   &  -1  \end{pmatrix}$.   In the noisy scenario, the parameter $\theta$ is affected by random fluctuations by an amount $\delta$, distributed according to a probability distribution $p(\delta)$.  The combined action of the signal and the noise is described by a single qubit dephasing channel
\begin{align}\label{qubitdephasing}
\map C_\theta (\rho)=& \int  \d\delta \, p(\delta )\,  U_{\theta+\delta }\, \rho \,  U_{\theta+\delta }^\dag ,
\end{align}
where $\rho$ is the initial density matrix of the system.

Traditionally, phase estimation is modelled as the task of estimating  the parameter $\theta$ with  $N$ independent accesses to  channel $\map C_\theta$.  In this setting, it is known that any finite  amount of dephasing noise compromises the Heisenberg scaling \cite{huelga1997improvement,fujiwara2008fibre,demkowicz2012elusive}, even if one adopts arbitrary error correction operations and fast quantum control \cite{sekatski2017quantum,zhou2021asymptotic}.   In many relevant scenarios, however, the channel $\map C_\theta$  only represents the effective evolution of a two-dimensional subspace of a larger quantum system. For example, a polarization qubit corresponds to the two-dimensional subspace  spanned by the states  $|{\rm H}\>  : = |1\>_{\rm H}\otimes |0\>_{\rm V}$ and $|{\rm V}\>  :  =|0\>_{\rm H}\otimes |1\>_{\rm V}$,  where the subscripts $\rm H$ and $\rm V$ refer to two modes of the electromagnetic field with vertical and horizontal polarization, respectively. In this picture, the qubit channel (\ref{qubitdephasing}) is just a restriction of the overall  dephasing channel
\begin{align}\label{fielddephasing}
\widetilde{\map C}_\theta ( \rho)=& \int \d\delta \, p(\delta )\,   \widetilde U_{\theta+\delta }\, \rho \,  \widetilde U_{\theta+\delta }^\dag\,  ,
\end{align}
where the unitary operator  $\widetilde U_\theta  =  \exp[-i  \theta  (  a^\dag a  -  b^\dag b)/2]$ represents the action of the phase shift  on the relevant modes of the electromagnetic field, and  $a$ ($b$) is the annihilation operator for the mode with  horizontal (vertical) polarisation. 

The full description of the dephasing process (\ref{fielddephasing})  allows one to analyze the situation where a single photon is sent through multiple paths, each path subject to an independent dephasing process. Each path is associated to two  polarization modes, say ${\rm H}_j$ and ${\rm V}_j$ for the $j$-th path, and a single photon in a superposition of multiple paths is associated to the Hilbert space spanned by Fock states with total photon number equal to 1. 
A convenient way to represent such states is to introduce a factorization of the Hilbert space in terms of   a path degree of freedom and   a polarization degree of freedom. A photon with polarization state $|\psi\>=  \alpha\,  | \rm H\>  +  \beta\, |\rm V\> $ placed on the $j$-th path is   represented by the state $|\psi\>  \otimes |j\>  :  =   (\alpha\,   |1\>_{{\rm H}_j}\otimes |0\>_{{\rm V}_{\rm j}}   +\beta\,  |0\>_{{\rm H}_j}\otimes |1\>_{{\rm V}_j}   )  \otimes   \prod_{k\not  =  j  }  |0\>_{{\rm H}_k}  \otimes |0\>_{{\rm V}_k}$.   A photon on a superposition of paths is then described by linear combinations of states of this form.

The situation where all the paths lead to independent dephasing processes is described by applying the channel  $\widetilde{\map C}_\theta$ on the modes associated to each path.  In particular, if the path degree of freedom is initialized in the maximally coherent state $|e_0\>=\sum_{j=0}^{M-1}|j\>/\sqrt M$,  the  state of the photon after traversing the paths is    \cite{chiribella2019quantum}
\begin{align}  
\nonumber \widetilde{ \map C}_\theta^{\otimes M}  \big(\rho\otimes |e_0\>\<e_0|\big)=&\frac{\map C_\theta(\rho)+ (M-1)F_\theta\rho F_\theta^\dag }{M}\otimes |e_0\>\<e_0|\\  &  +
\sum_{m=1}^{M-1} \frac{\map C_\theta(\rho)- F_\theta\rho F_\theta^\dag }{M}\otimes |e_m\>\<e_m| \,,   \label{superposition}
\end{align}
where   $\rho$ is  the initial density matrix of the polarization degree of freedom,    $\{|e_m\>\}_{m=0}^{M-1}$ is  the Fourier basis for the path degree of freedom, and  $F_\theta:  =  \int   \d\delta \, p(\delta )\,    U_{\theta  + \delta}  $.

By performing an interferometric measurement on the photon's path,  it is possible to separate cases in which the polarization is in the state  $\rho_+ \propto \map C_\theta(\rho)+ (M-1)F_\theta\rho F_\theta^\dag$ and cases in which the polarization is in the state $\rho_-\propto \map C_\theta(\rho)- F_\theta\rho F_\theta^\dag$. 
  The measurement can be implemented, e.g. by first applying the quantum Fourier transform on the modes, which can be implemented with a linear optical circuit \cite{reck1994experimental}. 
Note that the first term in Eq.~\eqref{superposition} converges to $F_\theta \rho  F_\theta^\dag$ in the large $M$ limit.  Direct calculation shows that $F_\theta$ is proportional to a unitary operator: explicitly, one has $F_\theta  =   |f(1)|  \,   U_{\theta +  \theta_0}$, where $f(k):=\int_{0}^{2\pi} {\d\delta } \, p(\delta )\,e^{-i  k  \delta /2}$ is the Fourier transform of the noise distribution, and  $\theta_0  =-2\, \arctan {\sf Im} [f(1)]  /{\sf Re} [f(1)]$.

Let us  consider the second term in Eq.~\eqref{superposition}.   In 
Appendix \ref{APP:Eq4}, 
we show that this term is proportional to a unitary phase shift if and only if  the Fourier transform of the noise distribution  satisfies the condition
\begin{align}\label{condition}
\left|f(2)-f^2\left(1 \right) \right| &=  1  -  \left|f^2\left(1 \right) \right| \,.
\end{align}
Under this condition, the  second term in Eq.~\eqref{superposition} is proportional to a phase shift by the amount $\theta  +  \theta_1$,  where $\theta_1$ is a fixed offset, depending on the noise distribution (see Appendix \ref{APP:Eq4}
for the exact expression). 

An  example of  noisy process  that satisfies condition (\ref{condition}) is a random phase kick that shifts the phase  by $\delta_0 $, where $\delta_0 \in  [0,2 \pi]$ is a fixed (but otherwise arbitrary)  offset.  In this model, the photon has probability $p$ to get a phase shift  $\theta$ and probability $1-p$ to get a phase shift $ \theta + \delta_0$, and $p$ can have any value between 0 and 1.      Physically, the random phase kick can be realized in a collisional model  for dephasing \cite{ziman2005all}, where it corresponds to the case of a qubit environment.     Moreover, the random phase kick model  is  important in that it corresponds to the short-time behavior induced by  the master equation  \cite{gardiner1991quantum} 
\begin{align}\label{continuous}\frac{\d \, \map C_{\omega,t } (\rho)}{\d t}  =-i\frac{\omega }{2} \left[Z, \map C_{\omega,t } (\rho)\right] + \frac{\gamma }{2}\left[ Z\,\map C_{\omega,t } (\rho)\, Z -\map C_{\omega,t }(\rho)\right] \, ,
\end{align}  
where $t$ is the time parameter, $\omega$ is the frequency, and $\gamma $ is the dephasing rate.  

When the condition (\ref{condition}) is satisfied, one can remove the offsets $\theta_0$ and $\theta_1$ in the two terms of Eq.~(\ref{superposition}), obtaining a noiseless channel in the limit  $M\to \infty$.  In the following, we analyze the  finite $M$ scenario, showing that Heisenberg limit can be achieved whenever $M$ grows linearly with $N$, where $N$ is the number of probes (for parallel protocols) or the number of time steps (for sequential protocols).   



\medskip
  
\begin{figure} 
\centering
\qquad \qquad \includegraphics[width=0.5\textwidth,trim=4 4 4 4,clip]{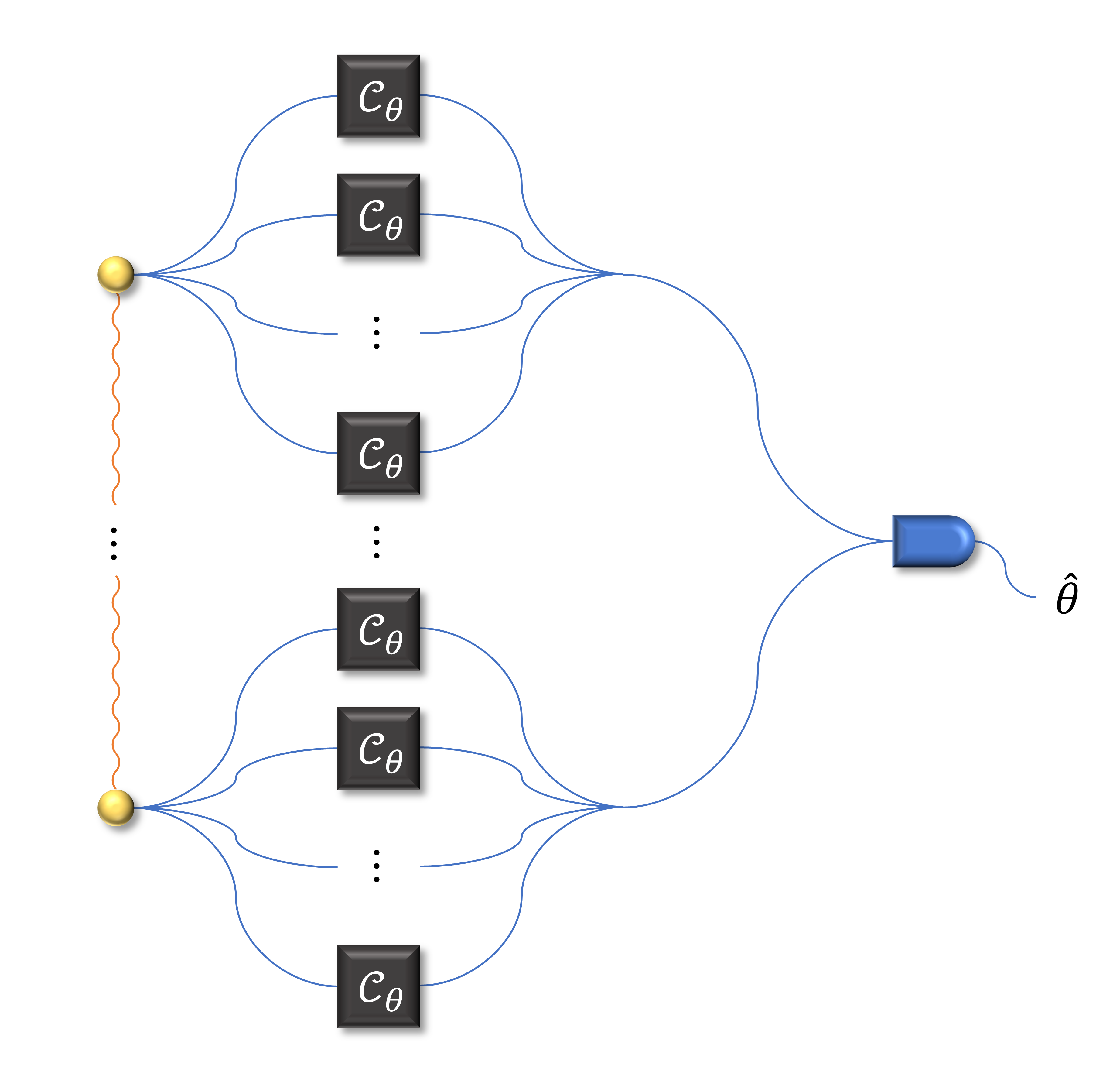}
\caption{{\bf Parallel protocol with coherent control on multiple probes.} 
 {  The unknown process $\map C_\theta$ is tested by  $N$ quantum probes, each of which propagates through  $M$ alternative paths.  The paths are then recombined to perform an interferometric measurement,  
 the outcomes of which are then used to produce an estimate of $\theta$.   }}\label{FIG:parallel}
\end{figure}

\emph{Parallel phase estimation protocol}.-- Let us start by considering the  scenario where the probe consists of $N$ entangled photons sent through  parallel uses of the same dephasing process, as shown in Fig. \ref{FIG:parallel}.

For simplicity, we will illustrate the ideas in the  basic  setting involving the preparation of the $N$ photons in a   GHZ state \cite{greenberger1989going}. 
 In the noiseless case, this state allow one to estimate small phase shifts in the interval $[0, 2\pi/N)$, and a setup using multiple GHZ states with different values of $N$ can achieve  Heisenberg scaling  of the error with respect to the total number of photons 
 \cite{xiang2011entanglement}. 
 We now provide a protocol that restores this ideal scaling in the presence of dephasing noise satisfying condition (\ref{condition}).  The steps of protocol are the  following: 
\begin{enumerate}
\item[(1)] Prepare $N$ photons in the polarization  entangled GHZ state $|\Psi_N\>=\left(|\text{H}\>^{\otimes N}+|\text{V}\>^{\otimes N}\right)/\sqrt 2$.
\item[(2)] Put each photon in a uniform superposition of $M$ paths, initializing the path degree of freedom in the maximally coherent state $|e_0\>$.
\item[(3)] Let the noisy process $\widetilde  {\cal C}_\theta$ act on each of the paths. 
\item[(4)] Perform a Fourier measurement on each path, getting outcome $m$. If the outcome is $m=0$, perform a phase shift of $-\theta_0$, otherwise perform a phase shift of $-\theta_1$.  
\item[(5)] Measure the polarization of the $N$ photons with a nonorthogonal measurement including the four operators $P_{\pm}  = |\Psi_{\pm}\>\<\Psi_\pm|/2$ and $Q_{\pm}  = |\Phi_{\pm}\>\<\Phi_\pm|/2$, with  
$|\Psi_\pm\>=(|\text{H}\>^{\otimes N}\pm |\text{V}\>^{\otimes N})/\sqrt 2$ and $|\Phi_\pm\>=(|\text{H}\>^{\otimes N}\pm  i\, |\text{V}\>^{\otimes N})/\sqrt 2$.
\item[(6)] Repeat the above procedure for $\nu$ times, and output the maximum likelihood estimate $\hat \theta = \arg \max_\theta \log p(x_1,\dots,x_\nu|\theta )$, where $\{x_1,\dots,  x_{\nu}\}$ are the outcomes of the $\nu$ measurements on the photons' polarization, and  $p(x_1,\dots, x_\nu |\theta )$ is the probability of obtaining such outcomes when  the true phase shift is $\theta$. 
\end{enumerate}

It is important to note that Step 4 can also be postponed to the end, and that the conditional phase shifts do not need to be implemented actively, as they can be included in the data processing stage.  However, our description of the protocol includes these operations because they simplify the presentation and analysis of the results.   

 In Appendix \ref{APP:Eq5}, 
 we  compute the  Fisher information for the outcomes of the measurement  at step (5), under the assumption that  Eq.  (\ref{condition}) is satisfied. Denoting the Fisher information by $F_\theta$, we prove the bound 
\begin{align}
F_\theta 
&\label{FI}\ge       \frac{N^2    \, \left|    1  -    \frac{ 1  -  e^{i\theta_0} f(2)  }M   \right|^{2N} }2     
\end{align}
This bound guarantees Heisenberg scaling whenever $M$ grows linearly with $N$. 

\emph{Sequential phase estimation protocol}.-- In this protocol,  a single photon is prepared in the polarization state $|+\>  =   ( |{\rm H}\>  +  |{\rm V}\>)/\sqrt 2$.   Then, the photon is sent in a uniform superposition of paths through $M$ independent dephasing channels. After the action of the channels, the paths are recombined, and a Fourier measurement is performed on the paths, followed by phase shifts that remove the offsets $\theta_0$ and $\theta_1$, in the same way as in the parallel protocol.  This procedure is repeated for $N$ steps, and a polarization measurement is finally performed after the $N$-th step, using a measurement  with operators $P_\pm  :=  |\pm\>\<\pm|/2 $ and $Q_{\pm}  =  |\pm i\>\<\pm i |/2$, with $|\pm\>  =  ( |{\rm H}\>  \pm  |{\rm V}\>)/\sqrt 2 $ and  $|\pm i\>  =  ( |{\rm H}\>  \pm  i  |{\rm V}\>)/\sqrt 2 $.   Also in this case, the intermediate measurements can in principle be postponed to the last step, and the conditional phase shifts of $\theta_0$ and $\theta_1$ can be absorbed into the data processing.   

In Appendix \ref{APP:Eq5}, 
we show that this sequential protocol is mathematically equivalent to the parallel one,  and that the Fisher information is still given by Eq. (\ref{FI}).    Hence, Heisenberg limit with respect to the number of time steps can be obtained with a single particle in a superposition of $M  =  O(N)$ paths per step.  

\begin{figure}[h]
\centering
 \includegraphics[width=0.52\textwidth,trim=4 6 4 4,clip]{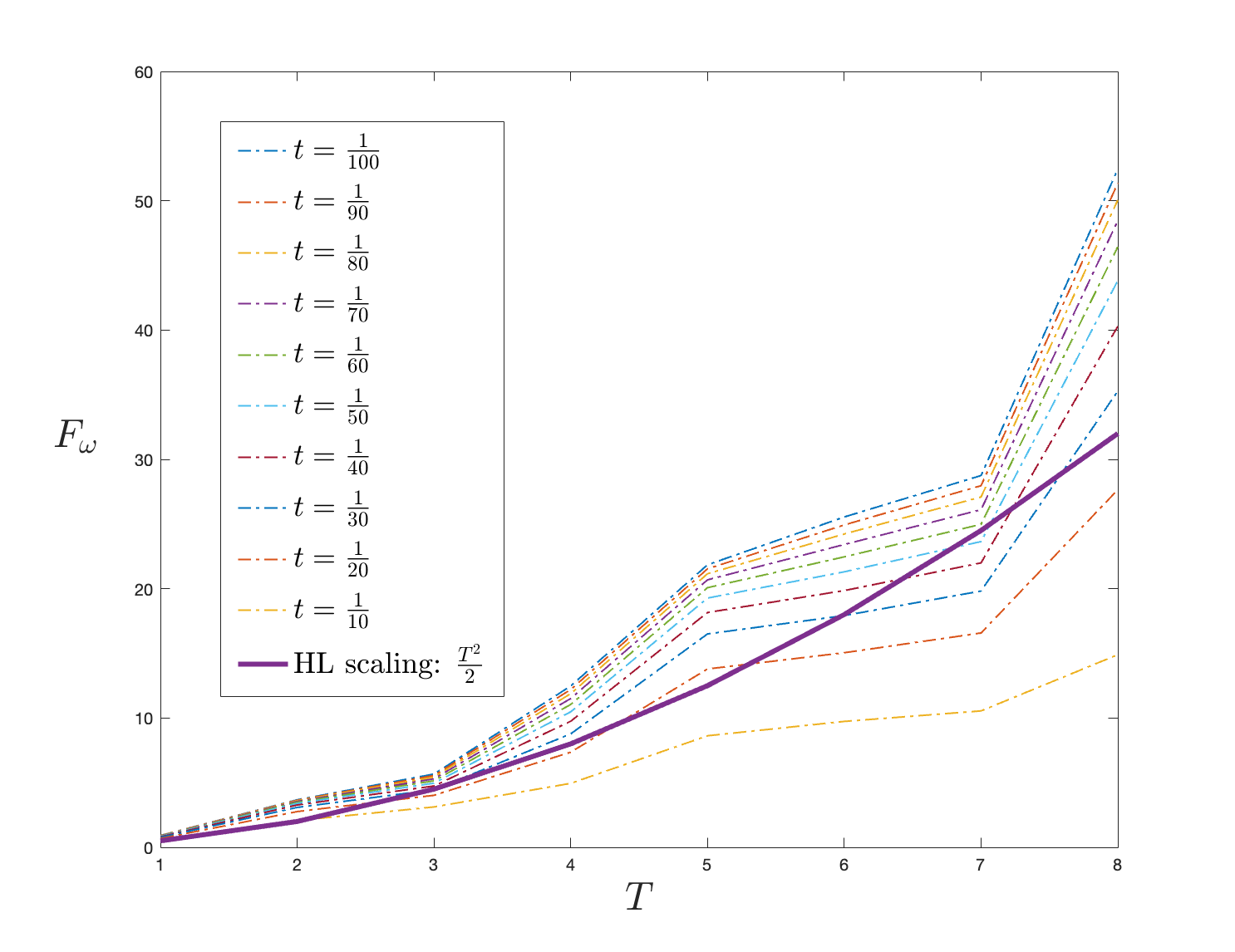}  
  \caption{{\bf Fisher information for frequency estimation with fast control operations.} 
    {The dotted lines show the Fisher information $F_\omega$ achieved by the sequential protocol with measurements applied at time steps of $t$ (measured in units where $\gamma=1$ (see Appendix \ref{APP:cont}
    for the details of the protocol).  The solid line  shows the asymptotic lower bound $F_\omega \ge   T^2/2$ for the Fisher information in the limit $t\to 0$.  In all plots we set   $\gamma=1$,   $\omega = \pi / 3$, and $M = T / t$. 
     }}\label{fig:continuous}
\end{figure}

The sequential protocol can also be applied to the estimation of the frequency $\omega$ in the master equation (\ref{continuous}). We consider the scenario where a single photon undergoes the evolution for a total time $T$, and fast control operations are applied at short intervals of time $t$  \cite{sekatski2017quantum}.  The control operations are measurements on the path degree of freedom, followed by appropriate shifts in the photon's polarization,  as in the discrete-time sequential protocol illustrated above.    In 
Appendix \ref{APP:cont}, 
 we show that the protocol achieves a Fisher information $F_\omega$ satisfying the bound  $F_\omega \ge T^2/2$ in the limit   $t\to 0$ using a number of paths growing as $T/t$.  The benefit of the superposition of paths carries over also to finite values of $t$, as illustrated  in Figure (\ref{fig:continuous}).

\emph{Conclusions. }--
In this paper we   explored the precision scaling achieved by  probing quantum processes in a coherent  superposition of configurations.   We  showed  that routing each probe on a superposition of  trajectories can unlock  Heisenberg scaling (with respect to number of probes,  total energy, or total probing time)  in the presence of dephasing noise induced by a random phase kick.  Our findings are in stark contrast with the scenario where the probes are sent on definite trajectories, in which case the random phase kick is known to prevent the Heisenberg scaling.

The key resource exploited by our protocols is quantum coherence in the probes' trajectories. Notably, this resource is different from other resources in quantum metrology, such as the total number/energy of the probes, or the total time.   The fact that the number of trajectories does not affect the total energy suggests that the superposition of trajectories could be used to achieve higher levels of precision in scenarios where the energy is bounded, such as e.g. in biological probes \cite{taylor2013biological}.       
   
Our  results have only scratched the surface of   the potential benefits of  the  superposition of configurations  in quantum metrology.  An interesting direction for future research is to consider superpositions of more complex configurations, for example including tree-like structures  in which the basic setups of this paper are used as building blocks. In this context, two key open problems arise. The first is to characterize what is the most general class of noise models for which the Heisenberg scaling can be restored through a superposition of configurations.  Recently developed  frameworks for  quantum circuits with quantum control  \cite{wechs2021quantum,vanrietvelde2021routed} are a valuable tool to address this problem.  The second problem is to extend the analysis to scenarios where the path degree of freedom is also subject to noise, e.g. due to imperfections of the beamsplitters used to route photons on different paths.   An interesting approach here is to consider concatenated  schemes, similar to those considered in quantum error correction and fault tolerance \cite{knill1996concatenated,boulant2005experimental,jochym2014using}.

Finally, an appealing direction is the experimental demonstration of quantum metrology boosted by coherent control on the probes' trajectories.  For moderate values of $N$,  the proof-of-principle demonstration of the protocols proposed appears to be within reach with current technologies of photonic quantum metrology \cite{barbieri2022optical,polino2020photonic}, especially in the sequential setting, which does not require multiphoton entanglement.

{\bf Acknowledgments}    This work was supported by the  Hong Kong Research Grant Council through grant  17300918 and though the Senior Research Fellowship Scheme SRFS2021-7S02, by the Croucher Foundation,  and by the John Templeton Foundation through grant  61466, The Quantum Information Structure of Spacetime  (qiss.fr).  Research at the Perimeter Institute is supported by the Government of Canada through the Department of Innovation, Science and Economic Development Canada and by the Province of Ontario through the Ministry of Research, Innovation and Science. The opinions expressed in this publication are those of the authors and do not necessarily reflect the views of the John Templeton Foundation.

\bibliographystyle{apsrev4-1}
\bibliography{reference}

\begin{widetext}

\appendix 

\section{Derivation of Eq. ~(4)}\label{APP:Eq4}
Here we show that the second term in Eq.~(3) 
is proportional to a unitary channel if and only if the condition in Eq.~(4)  
is satisfied.  By explicit evaluation, we have 
\begin{align}
\nonumber U_\theta^\dag \,  \map C_\theta  (\rho)\,  U_\theta   &  = \int  \d \delta \,p(\delta)  \, \left\{ \frac{ 1+  \cos \delta} 2  \,      \rho   +   \frac{ 1-  \cos \delta} 2 \,  Z\rho  Z \,       -   i   \frac{\sin \delta}2    Z  \rho   +     i   \frac{\sin \delta}2      \rho  Z         \right\}\\ 
&=  \frac{ 1+  {\sf Re} [f(2)] } 2  \,      \rho   +   \frac{ 1-  {\sf Re} [f(2)] } 2 \,  Z\rho  Z \,       +   i   \frac{  {\sf Im} [f(2)] }2    Z  \rho   -     i   \frac{ {\sf Im} [f(2)] }2      \rho  Z      \label{Cf2}
\end{align}
and 
\begin{align}
\nonumber U_\theta^\dag F_\theta  &=  \int  \d \delta \,  p(\delta)  \,  \cos \frac \delta 2  \,   I  -  i  \sin \frac \delta 2 \,  Z  \\
& =  {\sf Re}  [f(1)] \,  I    +i \,  {\sf Im}  [f(1)] \,  Z\, 
\end{align}
Combining these two relations, we obtain 
\begin{align}\label{secondterm}
U_\theta^\dag \,  \left(  \map C_\theta  (\rho)  -  F_\theta  \rho  F_\theta^\dag   \right)\,  U_\theta 
&  =  A_{00} \,      \rho   +     A_{11}\,  Z\rho  Z \,   + A_{01}  \,  Z \rho        +  A_{10}      \rho  Z  \, ,
\end{align}
where $(A_{ij})_{i,j\in  \{0,1\}}$ are the entries of the matrix
\begin{align}
A  = \begin{pmatrix}     \frac{ 1+  {\sf Re} [f(2)]   -  2 {\sf Re}  [f(1)]^2 }2    &     +   i   \frac{  {\sf Im} [f(2)]    - 2 {\sf Re}  [f(1)]  {\sf Im}  [f(1)]  }2      \\  
	-   i   \frac{  {\sf Im} [f(2)]    - 2 {\sf Re}  [f(1)]  {\sf Im}  [f(1)]  }2   &   \frac{ 1-  {\sf Re} [f(2)] -  2  {\sf Im}  [f(1)]^2}2 
\end{pmatrix} \, .
\end{align}
The map on the right hand side of Eq. (\ref{secondterm}) is proportional to a unitary channel if and only if the matrix $A$ has not full rank, that is, if and only if $\det (A)  =  0$.   Explicit calculation of the determinant yields 
\begin{align}
\det(A)  =  \frac{ \left|1-    |f(1)^2| \right|^2   -  \left|  f(1)^2  -  f(2)  \right|^2}4 \, .
\end{align}
Hence, we have $\det (A)  =  0$ if and only if $\left|1-    |f(1)^2| \right|   =  \left|  f(1)^2  -  f(2)  \right|$.   Using the inequality $|f(1)|  \le 1$, this condition can be rewritten as 
\begin{align}
1-    |f(1)^2|    =  \left|  f(1)^2  -  f(2)  \right| \, .
\end{align} This is the condition given in Eq. (4) of the main text.

When this condition is satisfied, one has 
\begin{align}\label{thetauno} 
U_\theta^\dag \,  \left(  \map C_\theta  -  F_\theta  \rho  F_\theta^\dag   \right)\,  U_\theta    =  (  1-  | f(1)^2 |)  ~  U_{\theta_1}  \,  \rho  U_{\theta_1}^\dag\, ,
\end{align}
with  $\theta_1  =   \arctan \{-i    A_{10}/[A_{00} -(1-|f(1)^2|)/2]\}  =  \arctan  \frac{2 {\sf Re}  [f(1)]  {\sf Im}  [f(1)] -{\sf Im} [f(2)]     }{ {\sf Re} [f(2)]   -   {\sf Re}  [f(1)]^2 + {\sf Im}  [f(1)]^2 }$.

\section{Derivation of Eq.~(5) in the main text} \label{APP:Eq5}

\subsection{Expression for the effective channel acting on the probe}\label{APP:postp}

In our protocol, the experimenter measures the path degree of freedom on the Fourier basis $\{|e_m\> \, , m=0,\cdots, M-1\}$, obtaining outcome $m$.  The experimenter performs a phase shift  $-\theta_0$  if  $m=0$, or a phase shift $-\theta_1$ if $m\not = 0$.    The effective evolution of the probe is given by the channel
\begin{align}
\map E_\theta (\rho)  &  =  \frac{ U_{\theta_0}^\dag\,   \left[ \map C_{\theta }   +  (M-1)\,  F_{\theta } \rho  F_{\theta}^\dag\right]  \,  U_{\theta_0}}M +   \frac{M-1}M ~   U_{\theta_1}^\dag  (\map C_\theta  -  F_\theta\rho F_\theta^\dag ) \,  U_{\theta_1} \,. 
\end{align}
Recall the relation $  U_{\theta_0}^\dag  F_\theta  =   |f(1)|  \,   U_{\theta }$, given in the main text, and   the relation $U_{\theta_1}^\dag  \left(  \map C_\theta  -  F_\theta  \rho  F_\theta^\dag   \right) U_{\theta_1}    =  (  1-  | f(1)^2 |)  ~  U_{\theta}   \rho  U_{\theta}^\dag$, valid when the condition $1-    |f(1)^2|    =  \left|  f(1)^2  -  f(2)  \right|$ is 
satisfied  [cf. Eq. (\ref{thetauno}) of this Appendix 
]. Using these two relations, we  obtain  
\begin{align}
\label{eq15}  \map E_\theta (\rho)&=\frac{1}{M} U_{-\theta_0}  \map C_\theta (\rho)U_{-\theta_0}^\dag+\frac{M-1}{M} U_\theta \rho U_\theta^\dag \,  .  
\end{align}

Now, Eq. (\ref{Cf2}) yields the relation $  \map C_\theta    (\rho) =   \begin{pmatrix}    \rho_{\rm HH}  &   \rho_{\rm HV}    \,   f(2)  \,  e^{-i\theta}   \\  
\rho_{\rm VH}    \, f(2)^*    \, e^{  i\theta}  &  \rho_{\rm VV}  \end{pmatrix}$, where $\rho_{ij}:=\<i|\rho|j\>,  \,  i, j \in  \{{\rm H}, {\rm V}\}$ are the matrix elements of $\rho$.
Using this relation,  Eq. (\ref{eq15}) becomes
\begin{align}
\nonumber \map E_\theta \left(\rho\right)=&\begin{pmatrix}
	\rho_{\rm HH} &\left(\frac{e^{i\theta_0} f(2) }{M} +\frac{M-1}{M}\right)e^{-i\theta }\rho_{\rm HV}\\
	\left(\frac{e^{-i\theta_0} f(2)^* }{M} +\frac{M-1}{M}\right)e^{i\theta }\rho_{\rm VH} &\rho_{\rm VV}
\end{pmatrix}\\
=&\begin{pmatrix}
	\rho_{\rm HH} &    \lambda \, e^{-i  \,(\theta   + \theta_2)  }\rho_{\rm HV}\\
	\lambda  \,e^{i  (\theta  +\theta_2) }\rho_{\rm VH} &\rho_{\rm VV}
\end{pmatrix}  \,,  \label{dephasing}
\end{align}
the second equality following from the definitions  $\lambda:  =  \left|    \frac{e^{i\theta_0} f(2) }{M} +\frac{M-1}{M}   \right|$ and $e^{-i  \theta_2}  :  = \left( \frac{e^{i\theta_0} f(2) }{M} +\frac{M-1}{M} \right) /  \lambda$. 

Equivalently, we have   $  \map E_\theta (\rho) =  \frac{1+\lambda}2 \,  U_{\theta+ \theta_2 }   \rho  U_{\theta+ \theta_2 }^\dag    +   \frac{1-\lambda}2 \,  U_{\theta+ \theta_2+\pi }   \rho  U_{\theta+ \theta_2 +\pi}^\dag$. 

\subsection{Achievable Fisher information in the parallel protocol}

In our first protocol, $N$ probes are initialized  in the entangled state  $|\Psi_+\>=\left(|\text{H}\>^{\otimes N}+|\text{V}\>^{\otimes N}\right)/\sqrt 2$,  and   undergo $N$ independent applications of the channel $\map E_\theta$ in Eq. (\ref{dephasing}). The resulting state is
\begin{align}
\map E_\theta^{\otimes N}\left(|\Psi_N^+\>\<\Psi^+_N | \right) =& \frac{1}{2} \left[  |\text{H}\>\<\text{H}|^{\otimes N}+|\text{V}\>\<\text{V}|^{\otimes N}   +\lambda^N e^{-iN(\theta+ \theta_2)}|\text{H}\>\<\text{V}|^{\otimes N}\right.\left.+\lambda^N e^{iN(\theta+ \theta_2)}|\text{V}\>\<\text{H}|^{\otimes N}\right]\,. \label{parallelE}
\end{align}
At this point, suppose that the experimenter implement a  measurement containing the operators $P_\pm  =   |\Psi_\pm\>\<\Psi_\pm|/2$ and $Q_{\pm}  = |\Phi_\pm\>\<\Phi_\pm|/2  $, with   $|\Psi_\pm\>=(|\text{H}\>^{\otimes N}\pm  |\text{V}\>^{\otimes N})/\sqrt 2$ and $|\Phi_\pm\>=(|\text{H}\>^{\otimes N}\pm  i  |\text{V}\>^{\otimes N})/\sqrt 2$. 
The probabilities of the corresponding outcomes are
\begin{align}\label{probmeas}
p(P_\pm|\theta)=\frac 1 4 \left[1\pm \lambda^N\cos (N(\theta+\theta_2))\right] \qquad{\rm and} \qquad p(Q_\pm|\theta)=\frac 1 4 \left[1\pm \lambda^N\sin (N(\theta+\theta_2))\right]
\end{align}

The corresponding classical Fisher information is: 
\begin{align}
F_\theta &\nonumber =\sum_{j  \in  \{  P_+,  P_-, Q_+,   Q_-\} } p(j|\theta) \left(\frac{\d \ln p(j|\theta)}{\d \theta}\right)^2\\
\nonumber &=  \frac{N^2    \, \lambda^{2N} }2  \,  \frac{1-  \lambda^{2N} + \frac{  \lambda^{2N}\,    \sin^2 [2N(\theta+\theta_2)] }2 }{1-\lambda^{2N}   + \frac{\lambda^{4N}  \,   \sin^2 [2N(\theta+\theta_2)}4 }\\
&  =   \frac{N^2    \, \lambda^{2N} }2  \, \left[ 1   +  \frac{    \frac{\lambda^{2N}\,    \sin^2 [2N(\theta+\theta_2)]}2  (1  -\lambda^{2N}/2  )}{1-\lambda^{2N}   + \frac{\lambda^{4N}  \,   \sin^2 [2N(\theta+\theta_2)}4 } \right]    \, . \label{FIpara}
\end{align}
From this exact  expression, we now derive  a  $\theta$-independent bound. Note that $\lambda  \le 1$: indeed, if we set $x:  =  {\sf Re}   [  e^{i\theta_0} f(2)]$ and  $y = {\sf Im}  [e^{i\theta_0} f(2)]$, we obtain 
\begin{align} 
\lambda   =\left|    \frac{e^{i\theta_0} f(2) }{M} +\frac{M-1}{M}   \right|  \le  \left|    \frac{e^{i\theta_0} f(2) }{M} \right|+\left|\frac{M-1}{M}   \right|   \le 1 \, ,
\end{align}
where we used the triangle inequality for the modulus, and  the relation $| f(2)|\le 1$. 

Since $\lambda \le 1$,  the second summand in Eq. (\ref{FIpara}) is nonnegative, and we have the bound   $F_\theta \ge    \frac{N^2    \, \lambda^{2N} }2 $, 
which coincides with  Eq. (5) in the main text.

\subsection{Achievable Fisher information in  the sequential protocol}
The sequential protocol amounts to $N$ repeated applications of the effective channel $\map E_\theta$ on the photon's polarization, initially in the state $|+\>$.  
The output state after the $N$-th application is 
\begin{align}
\map E_\theta^N (|+\>\<+|)    =      \frac 12  \,  \Big[  |{\rm H}\>\<{\rm H} |  +    |{\rm V}\>\<{\rm V} |   +    \lambda^{2N}    \,  e^{ - i  N  (\theta + \theta_2)}\,   |{\rm H}\>\<{\rm V} |    +   \lambda^{2N}    \,  e^{  i  N  (\theta + \theta_2)}\,   |{\rm V}\>\<{\rm H} |    \Big]\, .
\end{align} 
This equation is formally identical to Eq. (\ref{parallelE}).    At this point, the polarization undergoes  a measurement  with operators $P_\pm  :=  |\pm\>\<\pm|/2 $ and $Q_{\pm}  =  |\pm i\>\<\pm i |/2$, with $|\pm\>  =  ( |{\rm H}\>  \pm  |{\rm V}\>)/\sqrt 2 $ and  $|\pm i\>  =  ( |{\rm H}\>  \pm  i  |{\rm V}\>)/\sqrt 2 $. 
The outcome probabilities of this measurement coincide with the outcome probabilities in  Eq. (\ref{probmeas}), and therefore the Fisher information is still given by (\ref{FIpara}).

\section{Continuous-time dephasing}\label{APP:cont}

The single-qubit dynamics of a continuous-time Markovian dephasing  is characterized by the Lindblad master equation 
\begin{align}
\frac{\d \, \map C_{\omega,t } (\rho)}{\d t}=-i\frac{\omega }{2} \left[Z, \map C_{\omega,t } (\rho)\right] + \frac{\gamma }{2}\left[ U_\pi\,\map C_{\omega,t } (\rho)\, U_\pi^\dag  -\map C_{\omega,t }(\rho)\right]
\end{align}
where $\omega$ is the frequency of the oscillations, and $\gamma$ is the dephasing rate.    The solution of this equation is the quantum channel  $\map C_{\omega,t}( \rho)$ is the  channel $\map C_{\omega ,t}$ defined by 
\begin{align}\label{Comegat}
\map C_{\omega,t}( \rho) = \frac{1+e^{-\gamma  t}}{2}U_{\omega   t}\, \rho  \,U_{\omega   t}^\dag +\frac{1-e^{-\gamma  t}}{2}U_{\omega  t+\pi}\, \rho \, U_{\omega   t+\pi}^\dag
\end{align}  

When the qubit is the polarization of a single photon, the single-qubit dynamics (\ref{Comegat}) can be obtained from an extended  dynamics of the electromagnetic field, according to the master equation  
\begin{align}\label{eq22}
\frac{\d \, \widetilde{\map C}_{\omega,t } (\rho)}{\d t}=-i\frac{\omega }{2} \left[ a^\dag a-  b^\dag b, \widetilde{\map C}_{\omega,t } (\rho)\right] + \frac{\gamma }{2}\left[ \widetilde U_\pi  \,\widetilde{\map C}_{\omega,t } (\rho)\, \widetilde U_\pi^\dag -\widetilde{\map C}_{\omega,t }(\rho)\right] \, ,
\end{align}
where $a$ and $b$ are the annihilation operators for two modes of horizontal and vertical polarization, respectively, and $\widetilde U_\delta  :=    \exp  [-  i  (a^\dag a -  b^\dag b)  \delta/2]$ for arbitrary $\delta \in [0,2\pi]$.  

When the evolution is restricted to the subspace containing single-photon states and the vacuum,  the master equation  (\ref{eq22}) has the following solution: 
\begin{align}
\widetilde{\map C}_{\omega,t}(\tilde \rho)
=\left[ \begin{matrix}\rho_{\rm HH} & e^{-\gamma t-i\omega t }\, \rho_{\rm HV}&e^{-\frac{1+i}2\gamma t-i\frac \omega 2 t }\, \rho_{\rm Hvac} \\e^{-\gamma t+i\omega t }\, \rho_{\rm VH} &\rho_{\rm VV} &e^{-\frac{1-i}2\gamma t+i\frac \omega 2 t }\, \rho_{\rm Vvac}\\e^{-\frac{1-i}2\gamma t+i\frac \omega 2 t }\, \rho_{\rm vacH}&e^{-\frac{1+i}2\gamma t-i\frac \omega 2 t }\, \rho_{\rm vacV}&\rho_{\rm vacvac}\end{matrix}\right] \, . 
\end{align}
where $\tilde \rho \in \textbf{St}(\map H\oplus \map H_{\rm vac})$ is an arbitrary state in the space spanned by states $|\text{H}\>:=  |1\>_{\rm H} \otimes |0\>_{\rm V}$,    $|\text{V}\>:  =  |0\>_{\rm H} \otimes |1\>_{\rm V}$,  and  $|\text{vac}\>:=|0\>_{\text{H}}\otimes |0\>_{\text{V}}$, and $\rho_{ij}:=\<i|\tilde\rho|j\>, i,j\in \{\text{H},\text{V},\text{vac}\}$ are the matrix elements of $\tilde \rho$. This evolution can be rewritten in a compact way, as follows  
\begin{align}
\widetilde{\map C}_{\omega,t}(\tilde \rho)  =\map C_{\omega,t} (P \tilde \rho P) + F_{\omega,t} P \tilde \rho |\text{vac}\>\<\text{vac}|\label{eq4}+|\text{vac}\>\<\text{vac}|  \tilde \rho P F^\dag_{\omega,t} +  |\text{vac}\>\<\text{vac}|\tilde \rho|\text{vac}\>\<\text{vac}| , 
\end{align}
where $P$ is the projector on the original space $\map H=  \Span  \{  |\text{H}\> \, ,|\text{V}\>\}$,  and  $F_{\omega,t}:=e^{-\frac \gamma 2 t } U_{(\omega+\gamma) t }$.

Now, suppose that a single photon is sent on a superposition of $M$ paths, passing through $M$ independent instances of the channel $\widetilde{\map C}_{\omega , t}$. After the action of the channels, the paths are recombined and undergo a measurement on the Fourier basis. Depending on the outcome of the measurement, the photon's polarization is shifted either by $0$ (for outcome 0) or $-\theta_1$  (for outcomes other than 0), where $\theta_1$ will be defined later. The effective channel resulting from these operations is: 
\begin{align}
\nonumber \map E_{\omega,t} (\rho )&=\frac 1 M\left\{\map C_{\omega,t}(\rho)+(M-1)F_{\omega,t}\rho F_{\omega,t}^\dag+ (M-1) U_{-\theta_1}\left[\map C_{\omega,t}(\rho)-F_{\omega,t}\rho F_{\omega,t}^\dag\right]U^\dag_{-\theta_1}\right\} \\
&= \left[\begin{matrix}
	\rho_{\rm HH}&\lambda_t e^{-i\omega t}\rho_{\rm HV}\\
	\lambda_t^* e^{i\omega t }\rho_{\rm VH}&\rho_{\rm VV}
\end{matrix}\right] \, ,
\end{align}
with 
\begin{align}\label{lambdat}
\lambda_t  &=\frac{1}{M} e^{-\gamma t } +\frac{M-1}{M} e^{-\gamma(1+i) t  }+\frac{M-1}{M}e^{-\gamma t  +i \theta_1}-\frac{M-1}{M}e^{-\gamma(1+i) t +i\theta_1 } 
\end{align}

By applying the channel $\map E_{\omega,t}$   sequentially for $N=T/ t $ times, we then obtain   the channel 
\begin{align} \map E_{\omega,  t}^{ T/ t }     =\label{matrix1}
\left[\begin{matrix}
	\rho_{\rm HH}&\lambda_t^{T/ t } e^{-i\omega T}\rho_{\rm HV}\\
	\lambda_t^{*T/ t } e^{i\omega T }\rho_{\rm VH}&\rho_{\rm VV}
\end{matrix}\right]. 
\end{align}

We now consider  the problem of estimating the frequency $\omega$ for a given total time $T$ and a given dephasing rate $\gamma$.   For this purpose, we initialize the qubit in the state  $|+\>$  and we perform  a  measurement   with POVM operators $P_\pm  :=  |\pm\>\<\pm|/2 $ and $Q_{\pm}  =  |\pm i\>\<\pm i |/2$. The (classical) Fisher information achieved by this measurement  is 
\begin{align}
F_\omega&\nonumber =\sum_{j  \in  \{  P_+,  P_-, Q_+,   Q_-\} } p(j|\omega) \left(\frac{\d \ln p(j|\omega)}{\d \omega}\right)^2\\
& \label{eq28} =   \frac{T^2    \, |\lambda_{t}|^{2T/t} }2  \, \left[ 1   +  \frac{    \frac{|\lambda_{t}|^{2T/t}\,    \sin^2 [2 T(\omega+ \theta_t /t) ]}2  (1  -|\lambda_{t}|^{2T/t}/2  )}{1-|\lambda_{t}|^{2T/t}   + \frac{|\lambda_{t}|^{4T/t}  \,   \sin^2 [2 T(\omega+ \theta_t /t) ]}4 } \right]    , 
\end{align}
where the outcome  probabilities are $p(P_\pm|\omega)=[1\pm|\lambda_t|^{T/t}\cos (\omega T+\theta_tT/t)]/4$ and $p(Q_\pm|\omega)=[1\pm|\lambda_t|^{T/t}\sin (\omega T+\theta_tT/t)]/4$ with  $\theta_t$ being the phase  of the complex number $\lambda_t$.

We now show that the Fisher information has Heisenberg scaling in the limit $t\to 0$, corresponding to fast operations on the photon's path. 
For the  limit, we set  the number of paths $M$ to grow linearly with $T/t$, and optimize the choice of  $\theta_1$.  From Eq. (\ref{lambdat}) one can see that the maximum of $|\lambda_t|$ is obtained for 
\begin{align}
\theta_1  = \arg \left[  \frac{\frac{1}{M} e^{-\gamma t } +\frac{M-1}{M} e^{-\gamma(1+i) t  }}{    \frac{M-1}{M}e^{-\gamma t }-\frac{M-1}{M}e^{-\gamma(1+i) t  }   } \right] 
\end{align}
With this choice,  we  obtain 
\begin{align}
\nonumber \max_{\theta_1}\left|\lambda_t \right|& =e^{-\gamma t}\left(\left|\frac{1}{M}  +\frac{M-1}{M} e^{-i\gamma t  } \right|+\left|\frac{M-1}{M}\left(1-e^{-i\gamma t  }\right) \right|\right)\\
\nonumber &=e^{-\gamma t}\left(\left|1-i\gamma t +\mathcal O(t^2) \right|+\left|i\gamma t +\mathcal O(t^2)\right|\right)\\
\nonumber &=e^{-\gamma t }\left[1+\gamma t +\mathcal O(t^2)\right]\\
&=1-\mathcal O(t^2) \, .
\end{align}
To conclude, we  note that  Eq. (\ref{eq28}) implies the bound 
\begin{align}
F_\omega \ge \frac 1 2 T^2 \left|\lambda_t\right|^{2T/t} 
\end{align}
for sufficiently small $t$. 
Using the relation $\lim_{t\to 0}|\lambda_t|^{T/t}=1$, we then obtain  the asymptotic bound \begin{align}
\left.F_\omega \right|_{t\to 0}&\ge \frac 1 2 T^2 \, .
\end{align}

\end{widetext}

\end{document}